\documentstyle[prd,aps,preprint]{revtex}
\tightenlines
\input epsf.tex
\def\DESepsf(#1 width #2){\epsfxsize=#2 \epsfbox{#1}}
\def\Journal#1#2#3#4{{#1} {\bf #2}, #3 (#4)}

\def\NPB{{\em Nucl. Phys.} B}
\def\PLB{{\em Phys. Lett.}  B}
\def\PRL{\em Phys. Rev. Lett.}
\def\PRD{{\em Phys. Rev.} D}
\def\ZPC{{\em Z. Phys.} C}
\def\AP{\em Astropart. Phys.}
\def\PR{\em Phys. Rep.}
\def\IJMPA{{\em Int. J. Mod. Phys.} A}
\begin{document}

\draft
\preprint{\hbox{CTP-TAMU-05-01}}
\title{COANNIHILATION EFFECTS IN SUPERGRAVITY AND D-BRANE MODELS} 

\author{R. Arnowitt, B. Dutta and Y. Santoso }

\address{
Center For Theoretical Physics, Department of Physics, Texas A$\&$M
University, College Station TX 77843-4242}
\date{February, 2001} 
\maketitle
\begin{abstract}
Coannihilation effects in neutralino relic density calculations are
examined for a full range of supersymmetry parameters including large
$\tan\beta$ and large $A_0$ for stau, chargino, stop and sbottom coannihilation
with the neutralino. Supergravity models possessing grand unification with
universal soft breaking (mSUGRA), models with nonuniversal soft breaking in
the Higgs and third generation sparticles, and D-brane models with
nonuniversal gaugino masses were analysed. Unlike low $\tan\beta$ where $m_0$ is
generally small, stau coannihilation corridors with high $\tan\beta$ are highly
sensitive to $A_0$, and large $A_0$ allows $m_0$ to become as large as 1TeV.
Nonuniversal soft breaking models at high $\tan\beta$ also allow the opening of
a new annihilation channel through the s-channel $Z$ pole with acceptable
relic density, allowing a new wide band in the $m_0-m_{1/2}$ plane with
$m_{1/2} \stackrel{>}{\sim} 400$ GeV and $m_0$ rising to 1 TeV. The D-brane models considered possess stau
coannihilations regions similar to mSUGRA, as well as small regions of
chargino coannihilation. Neutralino-proton cross sections are analysed for
all models and it is found that future detectors for halo wimps will be
able to scan essentially the full parameter space with $m_{1/2} < 1$ TeV except
for a region with $\mu < 0$ where accidental cancellations occur when $5
\stackrel{<}{\sim}
\tan\beta \stackrel{<}{\sim} 30$. Analytic explanations of much of the above
phenomena are
given. The above analyses include current LEP bounds on the Higgs mass,
large $\tan\beta$ NLO correction to the $b \rightarrow s \gamma$ decay, and
large $\tan\beta$
SUSY corrections to the $b$ and $\tau$ masses.
\end{abstract}

\newpage
\section{Introduction}
Supersymmetry (SUSY) models with R-parity invariance offers a leading
candidate for the dark matter observed in the universe at large and locally
in the Milky Way galaxy. Thus in models of this type, the lightest
neutralino, $\tilde{\chi}_{1}^{0}$, is generally the lightest supersymmetric
particle (LSP)
and hence is absolutely stable. The observed dark matter is then the relic
neutralinos left over after the Big Bang. SUSY models are predictive since
they can calculate simultaneously the amount of relic density expected,
Milky Way detection signals and the production cross sections at
accelerators. Thus cosmological, astronomical and accelerator constraints
simultaneously constrain the SUSY parameter space.

The procedure for calculating the relic density of neutralinos is well
known. (For a review, see \cite{b1}.) Over the past decade, a number of
refinements of the analysis needed to get accurate answers have been
included. Thus the treatment of threshold and s-channel resonances in the
annihilation cross section in the early universe was discussed in
\cite{b2,b3,b4}.
In calculating the relic density, in most of the parameter space it is
sufficient to use the two body neutralino annihilation cross section, since
the effects of heavier particles are suppressed by the Boltzman factor.
However, in special situations, the next to lightest supersymmetric
particle (NLSP) may become nearly degenerate with the LSP, and the coupled
annihilation channels of LSP-LSP, NLSP-LSP, NLSP-NLSP must be
simultaneously considered~\cite{b2}. This phenomena of coannihilation was
considered
within the framework of the minimal supersymmetric model (MSSM) in \cite{b5} for
the situation when the lightest chargino, $\tilde{\chi}_{1}^{\pm}$, becomes nearly degenerate
with the $\tilde{\chi}_{1}^{0}$. In this paper we consider supergravity (SUGRA) models with
grand unification of the gauge coupling constants and soft SUSY breaking at
the GUT scale $M_{G} \cong 2 \times 10^{16}$ GeV. The low energy properties of the theory
is thus obtained by running the renormalization group equations (RGE) from
$M_{G}$ to the electroweak scale, where SUSY breaking at 
$M_{G}$ triggers $SU(2)_L\times U(1)_Y$
breaking at the electroweak scale. Aside from being in accord with the LEP
data implying grand unification, radiative breaking of $SU(2)_L\times U(1)_Y$ greatly
enhances the predictiveness of the theory as it reduces the number of SUSY parameters,
 and one finds generally for such
models that the $\tilde{\chi}_{1}^{0}$ is mostly Bino, and the 
$\tilde{\chi}_{1}^{\pm}$ is mostly Wino (when
the soft breaking masses are less than 1 TeV). Then the $\tilde{\chi}_{1}^{0}$
 and $\tilde{\chi}_{1}^{\pm}$ do
not become nearly degenerate and this type of coannihilation does not generally 
take place. (An exception for a class of D-brane models is discussed below.)

More recently it was pointed out that in SUGRA models, the sleptons
(particularly the lightest stau, $\tilde\tau_1$) can become nearly degenerate with
the $\tilde{\chi}_{1}^{0}$ leading to a new type of coannihilation, and this was explored
for low and intermediate values of $\tan\beta=\langle H_2 \rangle / \langle H_1
\rangle$~\cite{b6,b7}. (Here $H_{2,1}$
give rise to ($u$,$d$) quark masses). In this paper we examine this effect for
the full range of $\tan\beta$, i.e. $2 < \tan\beta < 50$~\cite{c12}. Further, in addition
to $\tilde{\tau}_1-\tilde{\chi}_1^0$ coannihilation, we find the possibility of
light stop $\tilde{t}_1-\tilde{\chi}_1^0$ coannihilation as well as light
sbottom $\tilde{b}_1-\tilde{\chi}_1^0$ coannihilation.

We consider these effects for a range of models
based on grand unfication of the gauge coupling constants: (1) Minimal
supergavity model (mSUGRA)~\cite{b8} where there is universal soft breaking masses
at $M_G$, (2) Nonuniversal supergravity models~\cite{b9}, where the first two generation
soft breaking masses are kept universal at $M_G$ (to suppress flavor changing
neutral currents) as well as the gaugino masses, but the Higgs and third
generation soft breaking masses are allowed to be nonuniversal, and (3) a
D-brane model~\cite{b10} based on Type IIB strings~\cite{b11} where soft breaking masses
at $M_G$ of the $SU(2)_L$ doublet squark, slepton and Higgs are kept degenerate
but distinct from the $SU(2)_L$ singlets, and similarly the gaugino 
 $SU(2)_L$ doublet soft breaking mass is distinct from the $SU(2)_L$
singlets. While the first two models have been characterized as SUGRA
models, they can also be realized in string models, and in fact any string
model based on grand unification at a high $M_G$ scale with the Standard Model
gauge group holding below $M_G$, phenomenologically can be treated as a SUGRA
model with an appropriate amount of nonuniversal soft breaking. Thus the
three models sample the possibilities of universal soft breaking,
nonuniversal soft breaking in the squark, slepton and Higgs sector but
universal gaugino masses, and finally nonuniversality in the gaugino sector
as well.

Since the Milky Way is perhaps 90$\%$ dark matter, it is a 
``laboratory" for studying the properties of dark matter (DM). Possible
signals for DM include annihilation in the halo of the Galaxy, annihilation
in the center of the Sun or Earth, and direct detection from scattering by
terrestial nuclear targets. Of these, the last is most promising, and we
restrict our discussion here to this. (For recent discussions of the other
possibilities see \cite{b12,b13,b14}). In general, the neutralino-nucleus
scattering has a spin independent and spin dependent part. However, for
heavy nuclei (aside from exceptional situations discussed below), the spin
independent scattering dominates, giving rise to approximately equal
scattering by neutrons ($n$) and protons ($p$) in the nucleus. It is thus
possible to extract from any data the neutralino-proton cross section,
$\sigma_{\tilde\chi^0_1-p}$ (subject to astronomical uncertainties about the Milky Way).
Current detectors (DAMA, CDMS, UKDMC) are sensitive to cross sections
\begin{equation}
\sigma_{\tilde\chi^0_1-p}\stackrel{>}{\sim} 1\times 10^{-6} \, {\rm pb}    
\label{eq1}
\end{equation}
with a one to two order of magnitude improvement possible in the near future.
Future detectors (e.g. GENIUS, Cryoarray) plan to be sensitive down to
\begin{equation}
\sigma_{\tilde\chi^0_1-p}\stackrel{>}{\sim} (10^{-9}-10^{-10}) \, {\rm pb}   
\label{eq2}
\end{equation}
and thus it is of interest to see what parts of the SUSY parameter space
can be examined by such detectors.

In order to obtain accurate calculations of both the relic density and 
$\sigma_{\tilde\chi^0_1-p}$
for large $\tan\beta$, it is necessary to include a number of corrections in
the analysis, and we list these here:
(i) In relating the theory at $M_G$ to phenomena at
the electroweak scale, the two loop gauge and one loop Yukawa renormalization group equations
(RGE) are used, iterating to get a consistent SUSY spectrum. (ii) QCD RGE corrections are
further included below the SUSY breaking scale for contributions involving light
quarks. (iii)
A careful analysis of the light Higgs mass $m_h$ is necessary (including two loop and pole
mass corrections) as the current LEP limits impact sensitively on the relic
density analysis. (iv) L-R mixing terms are included in the sfermion (mass)$^2$ matrices
since they produce important effects for large $\tan \beta$ in the third
generation. (v) One
loop corrections are included to
$m_b$ and
$m_{\tau}$ which are again important for large $\tan \beta$. (vi) The experimental bounds on the
$b\rightarrow s\gamma$ decay put significant constraints on the SUSY parameter space and
theoretical calculations here include the leading order (LO) and NLO corrections
for large $\tan \beta$.
Note that we have not in the following imposed
$b-\tau$ (or $t-b-\tau$) Yukawa unification or proton decay constraints as these depend
sensitively on unknown post-GUT physics. For example, such constraints do not naturally occur
in the string models where
$SU(5)$ (or $SO(10)$) gauge symmetry is broken by Wilson lines at $M_G$ (even though grand
unification of the gauge coupling constants at $M_G$ for such string models is
still required).

In carrying out the above calculations we have included the latest LEP
bound on the light Higgs mass, $m_h >$ 114 GeV~\cite{b15} and the recent NLO
corrections for large $\tan\beta$ for the $b\rightarrow s\gamma$
decay~\cite{b16,b17}. (We
have checked numerically that \cite{b16} and \cite{b17} give identical results.) Since
there are still some remaining errors in the theoretical calculation of $m_h$
as well as uncertainty in the top quark mass, $m_t =(175 \pm5)$ GeV, we will
conservatively assume here that $m_h >$ 110 GeV. In the MSSM, the constraint
on $m_h$ is $\tan\beta$ dependent as $Ah$ production ($A$ is the CP  odd Higgs) when
$m_A\stackrel{\sim}{=}M_Z$ is possible for $\tan\beta\stackrel{>}{\sim} 9$  and
this can be confused with $Zh$
production. However, in SUGRA models, radiative electroweak breaking
eliminates this part of the parameter space, and so it is correct to impose
the LEP bound on $m_h$ for the full range of $\tan\beta$. LEP also gives the
bound~\cite{b18} $m_{\tilde{\chi}_{1}^{\pm}} >$ 102 GeV, and the Tevatron bound
for the gluino $(\tilde g)$ is
\cite{b19} $m_{\tilde g} > $270 GeV (for squark ($\tilde q$) and gluino masses approximately
equal). We assume an allowed 2$\sigma$ range from the CLEO data for the 
$b\rightarrow s\gamma$ branching ratio~\cite{b20}:
\begin{equation} 
1.8\times 10^{-4}\leq B(B\rightarrow X_s\gamma)\leq 4.5 \times 10^{-4}   
\label{eq3}
\end{equation}

In calculating the relic density we will assume the bounds
\begin{equation}
               0.02 < \Omega_{\tilde\chi^0_1}h^2 < 0.25     \label{eq4}
\end{equation} 
where $\Omega_{\tilde\chi^0_1} = \rho_{\tilde\chi^0_1}/\rho_c$, 
$\rho_{\tilde\chi^0_1}$ 
is the relic density of the $\tilde\chi^0_1$, $\rho_c
= 3H_0^2/8\pi G_N$ ($H_0$ is the present Hubble constant and $G_N$ is the Newton constant)
and
$H_0 = h100$km\,s$^{-1}$Mpc$^{-1}$. The lower bound on 
$\Omega_{\tilde\chi^0_1}h^2$ is somewhat
lower than more conventional estimates 
($\Omega_{\tilde\chi^0_1} h^2  > 0.1$), and allows us
to consider the possibility that not all the DM are neutralinos. (In the
following, we will mention when results are sensitive to this lower bound.)
Accurate determinations by the MAP and Planck satellites of the dark matter relic density
will clearly strengthen the theoretical predictions, and already, analyses using combined data from the
CMB, large scale structure, and supernovae data suggests that the correct value of the relic
density lies in a relatively narrow band in the center of the region of
Eq.~(\ref{eq4})~\cite{b21}. We will
here, however, use the conservative range given in Eq.~(\ref{eq4}).

Supersymmetry theory allows one to calculate the $\tilde\chi^0_1$-quark cross section and we follow the
analysis of \cite{b22} to convert this to $\tilde\chi^0_1-p$ scattering. For this one needs
the $\pi N$ $\sigma$ term,
\begin{equation}
\sigma _{\pi N}={1\over 2}(m_u+m_d)\langle p|{\bar u} u+{\bar d}d|p\rangle,
\label{eq5}
\end{equation}
$\sigma_0=\sigma _{\pi N}-(m_u+m_d)\langle p|{\bar s} s|p\rangle$ and the quark mass ratio
$r=2 m_s/(m_u+m_d)$. We use here $\sigma_0= 30$ MeV
\cite{b23}, and $r=24.4\pm 1.5$\cite{b24}. Recent analyses, based on new $\pi-N$
scattering data gives $\sigma _{\pi N}\cong 65$ MeV\cite{b25}. Older 
$\pi-N$ data gave $\sigma _{\pi N}\cong 45$ MeV\cite{b26}.  We will use in most of the analysis
below the larger number. If the smaller number is used, it would have the overall effect in
most of the parameter space of reducing  $\sigma_{\tilde\chi^0_1-p}$ by about a
factor of 3. However, in
the special situation for $\mu < $0 where there is a cancellation of matrix elements, the
choice of $\sigma _{\pi N}$ produces a more subtle effect, and we will exhibit there results
from both values. Some of the results described below have been mentioned
earlier in conference talks~\cite{b27}.

There have been in the recent past a number of calculations of 
$\sigma_{\tilde\chi^0_1-p}$ in
the literature \cite{b23,b28,b29,b30,b31,b32,b33}. However, none include simultaneously all the
corrections discussed above, the large $\tan\beta$ range and the recent LEP $m_h$
bounds. We find general numerical agreement with other calculations for
regions of parameter space where the omitted corrections are small.

\section{ mSUGRA Model}

We consider in this section the mSUGRA model. This model is the most
predictive of the SUGRA models as it depends on only four additional
parameters and one sign. These may be chosen as follows: $m_0$, the universal
scalar soft breaking mass at $M_G$; $m_{1/2}$, the universal gaugino mass at 
$M_G$ (or
alternately one may use $m_{\tilde\chi^0_1}$ or $m_{\tilde g}$ as these quantities approximately scale
with $m_{1/2}$, i.e. 
$m_{\tilde\chi^0_1}\simeq 0.4 m_{1/2}$ and 
$m_{\tilde g}\simeq 2.8 m_{1/2}$); $A_0$, the universal
cubic soft breaking coupling at $M_G$; $\tan\beta = \langle H_2 \rangle /
\langle H_1 \rangle$; and $\mu/|\mu|$, the sign of the
Higgs mixing parameter $\mu$ in the superpotential term $W = \mu H_1H_2$. We
restrict the parameter space as follows:
\begin{eqnarray}
               m_0, m_{1/2} &<& 1 \,\,{\rm TeV} \label{eq6}\\
               |A_0/m_{1/2}| &<& 4 \label{eq7}\\
               2 < \tan\beta &<& 50 \label{eq8}
\end{eqnarray}
The relic density is calculated by solving the Boltzman equation in the
early universe. If the excited SUSY states lie significantly above the
$\tilde\chi^0_1$, then this takes the standard form~\cite{b1}:
\begin{equation}
 \frac{dn}{dt}=-3 H n- \langle \sigma_{ann} v \rangle
(n^2-n^2_{\rm eq})     \label{eq9}
\end{equation}
where $n$ is the neutralino number density, $n_{\rm eq}$ is its equilibrium
value, $\langle ... \rangle$ means thermal average, $\sigma_{\rm ann}$ is the annihilation
cross section ${\tilde\chi^0_1}
+ {\tilde\chi^0_1}\rightarrow f$ ($f$ = final state), and $v$ is the relative
velocity. In this case,
the effects of the heavier particles are suppressed by the Boltzman factor.
If however, one or more particles, $X_i$, have masses $m_i$ near the neutralino
($\tilde\chi^0_1 \equiv X_1$, $m_{\tilde\chi^0_1}\equiv m_1$), coannihilation
effects can occur. In this case Eq.~(\ref{eq9})
holds for the neutralino number density with, however, $n_{\rm eq}$ and
$\sigma_{\rm ann}$ given by~\cite{b2} 
\begin{equation}  
n_{\rm eq} = \sum_i n_{i{\rm eq}}     \label{eq10}
\end{equation}
\begin{equation} 
                 \sigma_{\rm ann} = \sum_{i,j}
		 r_ir_j\sigma_{ij};\,\,
		 r_i\equiv{n_{i{\rm eq}}\over n_{\rm eq}}    \label{eq11}
\end{equation}
In the nonrelativistic limit where freezeout occurs, the equilibrium number
densities take the form~\cite{b3,b34} 

\begin{equation}
     n_{i{\rm eq}}\cong
     g_i({{m_iT}\over{2\pi}})^{3/2}exp(-m_i/T)[1+{{15T}\over{8 m_i}}]  
     \label{eq12}
\end{equation}
Here $\sigma_{ij}$ is the cross section for $X_i +X_j \rightarrow f$
($\sigma_{11}$ is the neutralino annihilation cross section) and $g_i$ are the
degeneracy factors. In our analysis, we have included all final states that are
energetically allowed.

One sees from Eqs.~(\ref{eq11},\ref{eq12}) that the effects of coannihilation depend
simultaneously on the near degeneracy of any heavier particle with the
neutralino, and on the size of the  annihilation cross sections. A large cross section
can then compensate partially for the Boltzman factor. In SUGRA models two
effects occur which enhance the coannihilation for the sleptons. First,
since the neutralino is a Majorana particle, its annihilation cross section
is p-wave suppressed. Consequently,
\begin{equation}
         \sigma_{ij},\,\sigma_{i1}\stackrel{\sim}{=}O[10 \sigma_{11}];\,\,
	 i,j={\rm sleptons}    \label{eq13}
\end{equation}
and the slepton cross sections are enhanced relative to the neutralino
cross sections $\sigma_{11}$. Second, a numerical accident allows the selectrons to be
nearly degenerate with the neutralino over a region of parameter space of
increasing $m_0$, $m_{1/2}$. One can see this analytically for low and intermediate
$\tan\beta$ for mSUGRA where the RGE can be solved analytically~\cite{b35}. At the
electroweak scale one has
\begin{equation}
    \tilde m_{e_R}^2 = m_0^2  + (6/5) f_1 m_{1/2}^2 -\sin^2\theta_W M_W^2
    \cos(2\beta)     \label{eq14}
\end{equation}
\begin{equation}
   m_{\tilde{\chi}_{1}^{0}}= (\alpha_1/\alpha_G) m_{1/2}    \label{eq15}
\end{equation}
where $f_i = [1-(1+\beta_i t)^{-2}]/\beta_i$, $t= \ln(M_G/M_Z)^2$ and $\beta_1$ is the $U(1)$ $\beta$
function. Numerically this gives for e.g. $\tan\beta=$ 5
\begin{eqnarray}
          \tilde m_{\tilde e_R}^2 &=& m_0^2\, + \,0.15 m_{1/2}^2\,  + (37\,{\rm
	  GeV})^2 \nonumber \\    
             m_{\tilde{\chi}_{1}^{0}}^2 &=& 0.16 m_{1/2}^2      \label{eq16}
\end{eqnarray}
Thus for $m_0=$ 0, the $\tilde e_R$ becomes degenerate with the 
$\tilde{\chi}_{1}^{0}$ at $m_{1/2} = 370$ GeV,
i.e. coannihilation effects begin at $m_{1/2}\stackrel{\sim}{=}(350-400)$ GeV. For larger
$m_{1/2}$, the near degeneracy is maintained by increasing $m_0$, and there is a
corridor in the $m_0-m_{1/2}$ plane  allowing for an adequate relic density. 
(This is in accord with the numerical calculations of \cite{b29}.)

For larger $\tan\beta$, the situation is more complicated, as the the L-R
mixing in the $\tilde\tau_1$ (mass$^2$) matrix, generally makes the $\tilde\tau_1$ the lightest
slepton (lighter than the $\tilde e_R$ or $\tilde \mu_R$), and hence the $\tilde\tau_1$ dominates
the coannihilation phenomena. The RGE must of course now be solved
numerically, and there is strong sensitivity to $A_0$.
\begin{figure}[htb]
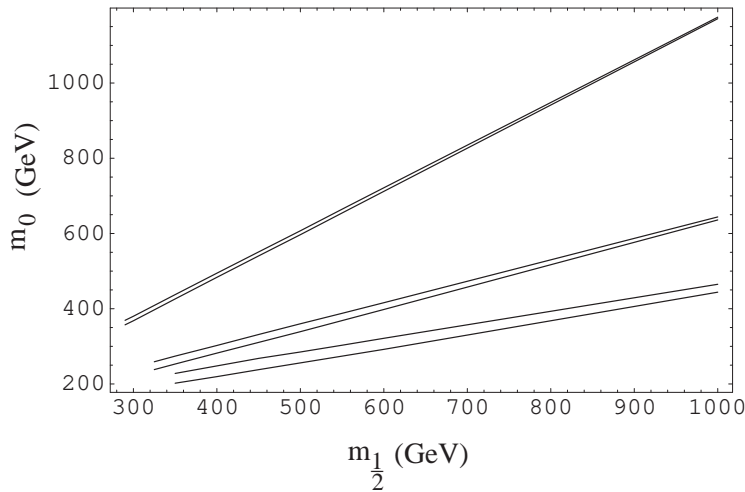

\centerline{ \DESepsf(aadcoan40new.epsf width 10 cm) }
\caption {\label{fig1} Allowed corridors in mSUGRA in the $m_0-m_{1/2}$ plane satisfying the
relic density constraint for $\mu > 0$, $\tan\beta = 40$ for (from bottom to top)
$A_0=m_{1/2}$, 2 $m_{1/2}$, 4$m_{1/2}$. The corridors terminate at low $m_{1/2}$ due to the 
$b\rightarrow s\gamma$ constraint. }
\end{figure}  
The new effects are
exhibited in Fig. 1 where the allowed regions in the $m_0-m_{1/2}$ plane
satisfying Eq.~(\ref{eq4}) for $A_0$=$m_{1/2}$, 2$m_{1/2}$ and 4$m_{1/2}$ are shown for $\mu > 0$ and
$\tan\beta = 40$. 
One sees several features of the large $\tan\beta$ region there.
First, the allowed corridors are sensitive to the value of $A_0$, the allowed values of $m_0$
increasing with $A_0$. Further, the
corridors end at a minimum value of $m_{1/2}$ due to the $b \rightarrow s\gamma$ 
constraint.
(We note that these lower bounds on $m_{1/2}$ are sensitive to the NLO
corrections to the $b\rightarrow s\gamma$  branching ratio mentioned in Sec. 1
above.) In fact, there is only the coannihilation region left in the
parameter space, since the $b \rightarrow s\gamma$ constraint at large $\tan\beta$ has
eliminated the non coannihilation part of the parameter space. (Actually,
there can be small islands of allowed parameter space left in the non
coannihilation region ($m_{1/2}\stackrel{<}{\sim}350$ GeV) for large $\tan\beta$ due to a
cancellation that can occur between the Standard Model amplitude and the
chargino amplitude.)  While for low $\tan\beta$, $m_0$ is generally
small~\cite{b29} ($m_0 \stackrel{<}{\sim}
200$ GeV), we see that at large $\tan\beta$, $m_0$ can become quite large reaching
up to 1 TeV for large $A_0$. Finally we mention that while for fixed $A_0$ the
allowed corridors are relatively narrow, as one allows $A_0$ to vary from
small to large values, the allowed region due to coannihilation covers most
of the $m_0-m_{1/2}$ plane.

We turn next to consider the ${\tilde{\chi}_{1}^{0}-p}$
 cross section. Large cross sections can
arise for large $\tan\beta$ and small $m_{\tilde{\chi}_{1}^{0}}$, and in
\cite{b32}, it
was shown that Eq.~(\ref{eq1}) 
could be satisfied in mSUGRA only also for small $\Omega_{\tilde{\chi}_{1}^{0}} h^2$, i. e. for
\begin{equation}
      \tan\beta \stackrel{>}{\sim}25,\,\,\,  
      \Omega_{\tilde{\chi}_{1}^{0}} h^2 \stackrel{<}{\sim} 0.1,\,\,\,
        m_{\tilde{\chi}_{1}^{0}}\stackrel{<}{\sim} 90\,{\rm GeV}    
	\label{eq17}
\end{equation}
Small cross sections generally arise for small $\tan\beta$, large $m_0$ and large
$m_{1/2}$. It is of interest to see what the minimum cross  section predicted by
theory is in order to know how much of the SUSY parameter space will be
examined by proposed detectors.  In order to see the effects of
coannihilation, we first examine the domain $m_{1/2} < 350$ GeV
($m_{\tilde{\chi}_{1}^{0}} < 140$ GeV)
where coannihilation does not take place. Fig. 2 exhibits the minimum cross
section for $\tan\beta = 6$, $\mu >0$, satisfying all the accelerator and relic
density constraints. One sees that 
$\sigma_{\tilde{\chi}_{1}^{0}-p} \stackrel{>}{\sim}1\times10^{-9}$ pb for $\tan\beta >
6$, and so the entire parameter space in this region would be accessible to
detectors planned with sensitivity of Eq.~(\ref{eq2}).
\begin{figure}[htb]
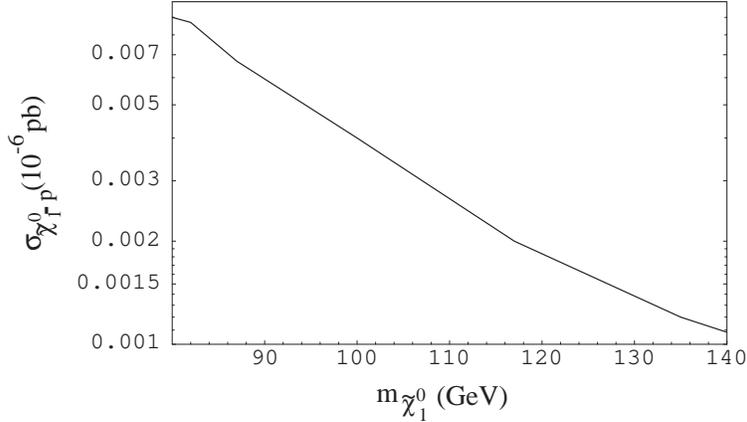

\centerline{ \DESepsf(aadtan6uninew.epsf width 10 cm) }
\caption {\label{fig2} $(\sigma_{\tilde{\chi}_{1}^{0}-p})_{\rm min}$ in the noncoannilation region as a function of 
$m_{\tilde{\chi}_{1}^{0}}$ for
mSUGRA for $\mu >0$, $\tan\beta = 6$, obtained by varying over the range of $m_0$ and
$A_0$.}
\end{figure}

At higher $m_{1/2}$, coannihilation occurs, allowing $m_0$ and $m_{1/2}$ to become
larger, and hence the cross section should become smaller. We examine first
the case $\mu >0$. The $A_0$ dependence at large $\tan\beta$ shown in Fig. 1 then
implies that the cross section  should also decrease with increasing $A_0$,
since, as seen in Fig. 1, $m_0$ then increases. This is seen explicitly in
Fig. 3. 
\begin{figure}[htb]
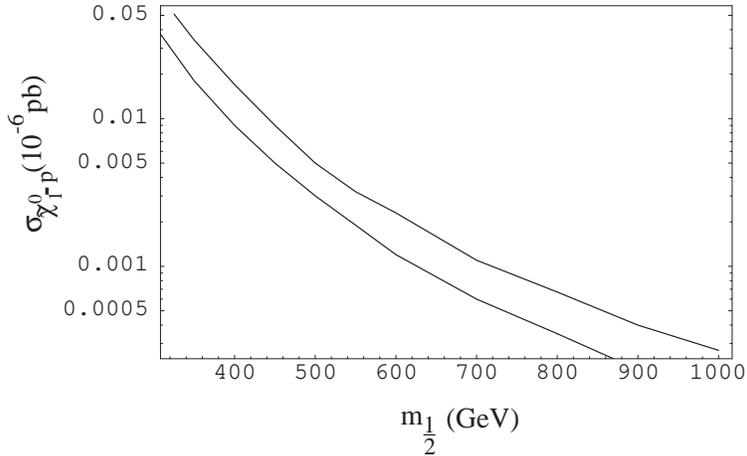

\centerline{ \DESepsf(aadcoan4024new.epsf width 10 cm) }
\caption {\label{fig3}$\sigma_{\tilde{\chi}_{1}^{0}-p}$
 as a function of $m_{1/2}$ for mSUGRA $\mu >0$, $\tan\beta=40$ and
$A_0=2 m_{1/2}$ (upper curve), $4 m_{1/2}$(lower curve).
 }
\end{figure}
Thus it is possible that as $m_{1/2}$ gets large, a consequence of coannihilation is that large $\tan\beta$ and
large $A_0$ can give smaller cross sections than low $\tan\beta$. This is
illustrated in Fig. 4 where $\sigma_{\tilde{\chi}_{1}^{0}-p}$
 is plotted as a function of $m_{1/2}$ for
$A_0=4\,m_{1/2}$, $\tan\beta=$ 40 (upper curve) and $\tan\beta = 3$ (lower
curve). One sees
that for $m_{1/2}>600$ GeV, the $\tan\beta = 40$ curve actually drops below
the $\tan\beta = 3$ curve, the $A_0$ dependence (producing large $m_0$) compensating
for the $\tan\beta$ dependence. One has however, the bound
\begin{equation}
          \sigma_{\tilde{\chi}_{1}^{0}-p}\stackrel{>}{\sim}1\times 10^{-10}\,\, {\rm pb}, 
	  \,\,\, {\rm for} \,\,\, \mu > 0,\,\, m_{1/2} < 1\,\, {\rm TeV}   
	  \label{eq18}
\end{equation}
Again, most of this parameter space will be within the reach of planned future
detectors.
\begin{figure}[htb]
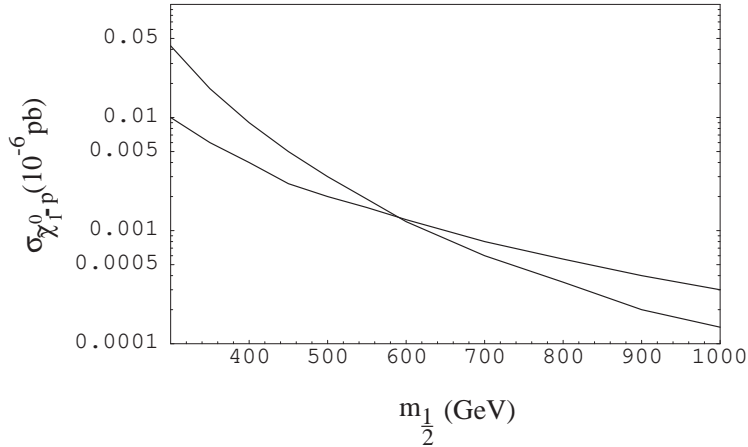

\centerline{ \DESepsf(aadcoan4034new.epsf width 10 cm) }
\caption {\label{fig4}$\sigma_{\tilde{\chi}_{1}^{0}-p}$ as a function of $m_{1/2}$ 
for mSUGRA, $\mu > 0$ for $A_0 = 4\,m_{1/2}$,
$\tan\beta = 40$ (upper left curve), $\tan\beta = 3$ (lower left curve).}
\end{figure}

We turn next to the case of $\mu < 0$. The situation here is more complicated.
As pointed out for low and intermediate $\tan\beta$ in \cite{b30}, an accidental
cancellation can occur between the Higgs $u$-quark and the Higgs $d$-quark
scattering 
amplitudes which can greatly reduce 
$\sigma_{\tilde{\chi}_{1}^{0}-p}$ and we
investigate here whether this cancellation continues to occur in the high
$\tan\beta$ region. Since the cancellation is somewhat subtle, we briefly
describe how it comes about. In general, $\tilde\chi^0_1$, is a mixture of
Bino($\tilde B$), Wino($\tilde W$) and the two Higgsinos ($\tilde H_1,$ $\tilde
H_2$):
\begin{equation}
\tilde\chi^0_1=N_{11}\tilde B+N_{12}\tilde W+N_{13}\tilde H_1+
N_{14}\tilde H_2       \label{eq19}
\end{equation}
where $N_{1i}$ are the amplitudes. In much of the parameter space, the t-channel
Higgs exchanges ($h$, $H$) dominate $\sigma_{\tilde{\chi}_{1}^{0}-p}$. 
The $d$ and $u$ quark Higgs amplitudes are~\cite{b37}
\begin{eqnarray}
A^d = \frac{g_2^2 m_d}{2 M_W} \left( - \frac{\sin \alpha}{\cos \beta}
\frac{F_h}{m_h^2} + \frac{\cos \alpha}{\cos \beta} \frac{F_H}{m_H^2} \right)
\label{eq20} \\
A^u = \frac{g_2^2 m_u}{2 M_W} \left( \frac{\cos \alpha}{\sin \beta}
\frac{F_h}{m_h^2} + \frac{\sin \alpha}{\sin \beta} \frac{F_H}{m_H^2} \right)
\label{eq21}
\end{eqnarray}
where $\alpha$ is the Higgs mixing angle and
\begin{eqnarray}
F_h = (N_{12}-N_{11} \tan \theta_W)(N_{14} \cos{\alpha} + N_{13} \sin{\alpha})
\label{eq22} \\
F_H = (N_{12}-N_{11} \tan \theta_W)(N_{14} \sin{\alpha} - N_{13} \cos{\alpha})
\label{eq23}
\end{eqnarray}
Since the
$s$-quark contribution to the scattering is quite large, the $d$-quark amplitude
$A^d$ will generally be quite large. However $A^d$ will be suppressed if the
amplitudes $N_{13}$, $N_{14}$ obey the equation
\begin{equation}
N_{14}\simeq-N_{13}\frac{\tan\alpha+{m_h^2\over m_H^2}\cot\alpha}{1+{m_h^2\over
m_H^2}} \label{eq24}
\end{equation}
In general $\tan\alpha$ is negative and
small ($\tan\alpha \stackrel{\sim}{=} - \cot \beta \simeq -0.1$), and
Eq.~(\ref{eq24}) can happen if $N_{14}/N_{13}$ is
positive, which is the case for $\mu < 0$. Once $A^d$ is sufficiently suppressed, it can cancel the remaining
(smaller) $u$-quark contribution coming from $A^u$, leading to a nearly zero value of
$\sigma_{\tilde{\chi}_{1}^{0}-p}$.
For fixed $\tan\beta$, total cancellation occurs at a fixed value of $m_{1/2}$,
though the effects of cancellation has a width.
What happens is shown in Fig. 5 where 
$\sigma_{\tilde{\chi}_{1}^{0}-p}$ is plotted
in the large $m_{1/2}$ region for $\tan\beta = 6$ (short dash), 
$\tan\beta = 8$
(dotted), $\tan\beta = 10$ (solid), $\tan\beta = 20$ (dot-dash) and 
$\tan\beta = 25$
(dashed). One sees that the cross section dips sharply when $\tan\beta$ is
increased from 6 to 8, and  and goes through a
minimum for $\tan\beta = 8$ at $m_{1/2}\stackrel{\sim}{=} 810$ GeV. For
$\tan\beta = 10$, the minimum
recedes to $m_{1/2}\stackrel{\sim}{=} 725$ GeV, and then begins to advance for higher 
$\tan\beta$
i.e. rising to $m_{1/2}\stackrel{\sim}{=}830$ GeV for $\tan\beta=$ 20, and 
$m_{1/2}\stackrel{\sim}{=}950$ GeV for
$\tan\beta = 25$. Thus if we restrict $m_{1/2}$ to be below 1 TeV, the cross section
will fall below the sensitivity of the planned future detectors for $m_{1/2}
\stackrel{>}{\sim}450$ GeV for a restricted region of $\tan\beta$ i. e.
\begin{equation}
  \sigma_{\tilde{\chi}_{1}^{0}-p} < 1\times10^{-10}\,\,
   {\rm pb} \,\,\,\,
   {\rm for}\,\,\,\, 450\,
   {\rm GeV} < m_{1/2} < 1\,{\rm TeV};\,\, 
   5 \stackrel{<}{\sim} \tan\beta\stackrel{<}{\sim}30    \label{eq25}
\end{equation}
\begin{figure}[htb]
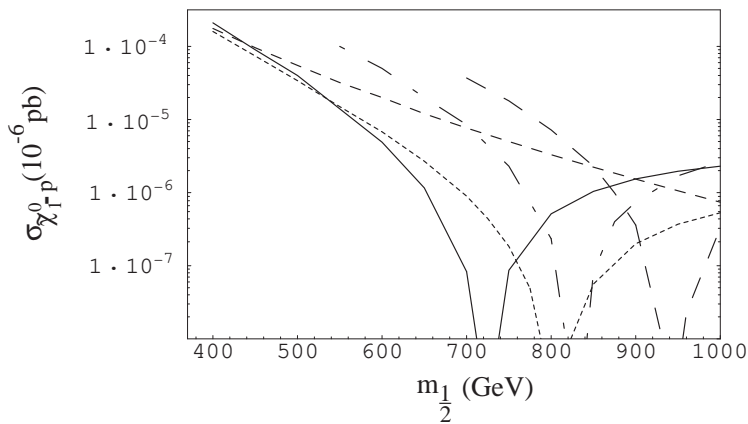

\centerline{ \DESepsf(aadcoan61020.epsf width 10 cm) }
\caption {\label{fig5} $\sigma_{\tilde{\chi}_{1}^{0}-p}$
 for mSUGRA for $\mu < 0$, $A_0 = 1500$ GeV, for $\tan\beta = 6$
(short dash), 
$\tan\beta = 8$ (dotted), $\tan\beta = 10$ (solid), $\tan\beta = 20$
(dot-dash), $\tan\beta=25$ (dashed). Note that the $\tan\beta = 6$ curve terminates
at low $m_{1/2}$ due to the Higgs mass constraint, and the other curves
terminate at low $m_{1/2}$ due to the $b \rightarrow s\gamma$ constraint.}
\end{figure}

At the minimum, the cross sections can become quite small, e. g. $<
1\times10^{-13}$ pb, without major fine tuning of parameters, corresponding to an
almost total cancellation. (Actually, as pointed out in \cite{b38}, the true minimum
of the cross section would then arise from the spin dependent part of
the scattering, which though very small, does not possess the same
cancelation phenomena.) Further, the widths of the minima for fixed
$\tan\beta$ are fairly broad. While in this domain the proposed detectors obeying
Eq.~(\ref{eq2}) will not be able to observe Milky Way wimps, mSUGRA would imply
that the squarks and gluino would lie above 1 TeV, but at masses that would
still be accessible to the LHC (i.e. $<2.5$ TeV). Also note that this phenomena occurs only
for $\mu < 0$ and for a restricted range of $\tan\beta$, so
cross checks of the theory would still be available.
	   
The minima occurring in Fig. 5 arose from cancellations between the Higgs
$u$-quark and Higgs $d$-quark amplitudes, and as such are sensitive to the
quark content of the proton. As discussed in Sec. 1, $\sigma_{\pi N}$ is not yet
well determined, and Fig. 5 used $\sigma_{\pi N} = 65$ MeV. In Fig. 6, we compare
this choice(solid curve)  for $\tan\beta$ = 10 with the parameters used in
\cite{b26,b30}(dashed curve),
where $\sigma_{\pi N}=$ 45 MeV. One sees that the
minimum at $m_{1/2} =$ 725 GeV is shifted to $m_{1/2}=$ 600 GeV, with analagous
shifts occurring for the other values of $\tan\beta$. Thus the extreme
cancellations that occur in the coannihilation region for $\mu <$ 0 are quite
sensitive to the properties of the proton.
\begin{figure}[htb]
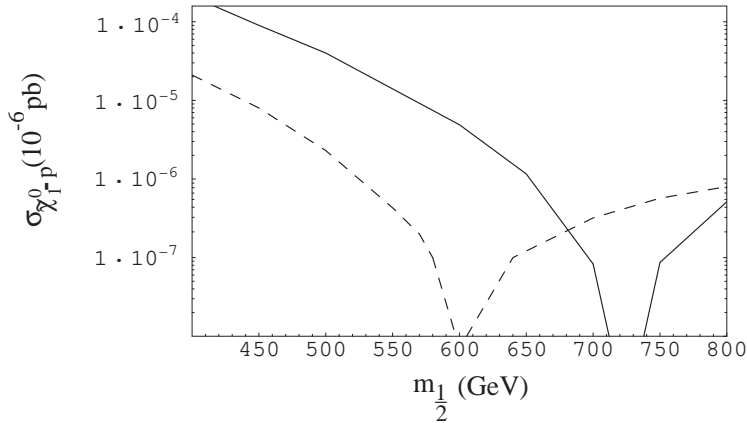

\centerline{ \DESepsf(aadcoanellis10.epsf width 10 cm) }
\caption {\label{fig6} $\sigma_{\tilde{\chi}_{1}^{0}-p}$ for mSUGRA for $\mu <$
0, $\tan\beta =10$, $A_0 =$ 1500 GeV for
$\sigma_{\pi N}=$ 65 MeV (solid curve) and the parameters of [26,36] (dashed curve)
where $\sigma_{\pi N}=$ 45 MeV was used.}
\end{figure}

\section{SUGRA Models With Nonuniversal Soft breaking}

We consider here models with nonuniversal soft breaking masses. In order to
suppress flavor changing neutral currents, we maintain the universal soft
breaking slepton and squark masses, $m_0$, at $M_G$ in the first two generations.
However, experiment is compatible with nonuniversal Higgs and third
generation soft breaking at $M_G$, and these can arise from nonuniversal
couplings in the supergravity Kahler potential. We parameterize this situation as
follows:
\begin{eqnarray} 
m_{H_{1}}^{\ 2}&=&m_{0}^{2}(1+\delta_{1}); 
\quad m_{H_{2}}^{\ 2}=m_{0}^{2}(1+ \delta_{2});\nonumber \\ m_{q_{L}}^{\
2}&=&m_{0}^{2}(1+\delta_{3}); \quad m_{t_{R}}^{\ 2}=m_{0}^{2}(1+\delta_{4});
\quad m_{\tau_{R}}^{\ 2}=m_{0}^{2}(1+\delta_{5});  \nonumber \\ m_{b_{R}}^{\
2}&=&m_{0}^{2}(1+\delta_{6}); \quad m_{l_{L}}^{\ 2}=m_{0}^{2}(1+\delta_{7}).
\label{eq26}
\end{eqnarray}
where $q_L = (\tilde t_L, \tilde b_L)$ squarks, $l_L = (\tilde \nu_\tau, 
\tilde \tau_L)$ sleptons etc. The
$\delta_i$ measure the deviations from universality. (If one where to impose
$SU(5)$ or $SO(10)$ symmetry then $\delta_3 = \delta_4 = \delta_5 \equiv
\delta_{10}$, and
$\delta_6 = \delta_7 \equiv \delta_{\bar5}$.) In the following we limit the $\delta_i$ to
obey
\begin{equation}
                       -1 < \delta_i < +1      \label{eq27}
\end{equation}
The lower bound is necessary to prevent tachyons at $M_G$, and the upper bound
represents a naturalness condition. We maintain gauge coupling constant
unification at $M_G$ (which is in accord with LEP data). We also assume here
gaugino mass unification. (Deviations from this can arise from nonuniversal
couplings in the supergravity gauge function $f_{ab}$, and an example of this is
treated in the next section.)

While the nonuniversal models contain a number of additional parameters,
one can obtain an understanding of the new effects these imply from the
following analytical considerations. In the decomposition of Eq.~(\ref{eq19}),
the lightest neutralino is mostly Bino, i.e. $N_{11}\stackrel{>}{\sim}0.8$. The Higgs mixing parameter $\mu$ to a large extent
controls this mix, and as $\mu^2$ decreases (increases) the higgsino content
increases (decreases). In order to see qualitatively what happens we examine the
 low and intermediate $\tan\beta$ region where the RGE can be
solved analytically. One finds for $\mu^2$ the result (see e.g. \cite{b39}):
\begin{eqnarray}
\mu^2&=&{t^2\over{t^2-1}}\left[({{1-3 D_0}\over 2}+{1\over
t^2})+{{1-D_0}\over2}(\delta_3+\delta_4)\right. \nonumber \\ 
&-&\left.{{1+D_0}\over2}\delta_2+{\delta_1\over
t^2}\right]m_0^2+{\rm {universal\,parts\,+\,loop \, corrections}}. \label{eq28}
\end{eqnarray}
where $t = \tan\beta$, and 
$D_0 \stackrel{\sim}{=} 1 - (m_t/200 {\rm GeV} \sin\beta)^2 \stackrel{\sim}{=} 0.25$. One sees that the
universal contribution to the $m_0^2$ term in $\mu^2$ is small, making $\mu^2$
quite sensitive to nonuniversal contributions. Also
since $D_0$ is small, the Higgs and squark nonuniversalities enter
coherrently, approximately in the combination $\delta_3 + \delta_4 -\delta_2$.
(The RGE can also be solved analytically for the SO(10) model where $\tan\beta >
40$, and numerically otherwise, with results qualitatively similar to
Eq.~(\ref{eq28}).)
In addition we note here the mass of the CP odd Higgs boson, $m_A$ is
\begin{eqnarray}
m_A^2&=&{{t^2+1}\over{t^2-1}}\left[{{3(1- D_0)}\over
2}+{{1-D_0}\over2}(\delta_3+\delta_4)\right. \nonumber  \\
&-&\left.{{1+D_0}\over2}\delta_2+\delta_1\right]m_0^2+{\rm
{universal\,parts\,+\,loop \, corrections}}.   \label{eq29}
\end{eqnarray}   
The effects of the nonuniversalities can be seen to fall into the following
categories:

(i)   The spin independent neutralino-proton cross section depends upon the
interference between the gaugino and higgsino parts of the neutralino, the
larger the amount of interference the larger the cross section. Thus
$\sigma_{\tilde{\chi}_{1}^{0}-p}$ can be significantly increased relative to
mSUGRA by decreasing $\mu^2$ (and hence increasing the higgsino content) by
chosing $\delta_3$, $\delta_4$, $\delta_1 < 0$, $\delta_2 >0$, and be decreased
(though by not so large amount) with the opposite sign choice.

(ii)  In the early universe annihilation cross section, the s-channel $Z^0$
pole amplitude depends on the higgsino contribution of the neutralino
(the Bino and Wino parts giving zero coupling). Thus if $\mu^2$ is lowered by chosing
$\delta_3$, $\delta_4$, $\delta_1 < 0$, $\delta_2 > 0$ one gets an increased amount of
annihilation, and a decreased amount with the opposite choice of signs.
Thus deviations from universality can significantly effect the satisfaction
of the relic density constraints of Eq.~(\ref{eq4}).

(iii) The third generation squark and slepton nonuniversalities shift the
mass spectrum of the particles. Thus the positions of the corridors of
coannihilation, and their widths can be significantly modified (also
modifying the cross section $\sigma_{\tilde{\chi}_{1}^{0}-p}$.)

Effects of type (i) were early observed in \cite{b39}, and further discussed in
\cite{b32}
where is was seen that $\sigma_{\tilde{\chi}_{1}^{0}-p}$ could be increased by a factor of 10 or
more with the sign choice $\delta_{3,4,1} < 0$, $\delta_2 > 0$. We here update this
result to take into account of the latest LEP bound $m_h > 114$ GeV. Fig. 7
shows the maximum value of $\sigma_{\tilde{\chi}_{1}^{0}-p}$ for the nonuniversal models for $\mu > 0$
with the above choice of signs for $\delta_i$ for $\tan\beta = 7$ (lower curve)
and $\tan\beta =12$ (upper curve). One sees that detectors obeying
Eq.~(\ref{eq1}) are
sampling the parameter space with $\tan\beta \geq 7$ (compared with the mSUGRA
result of $\tan\beta \geq 25$).

\begin{figure}[htb]
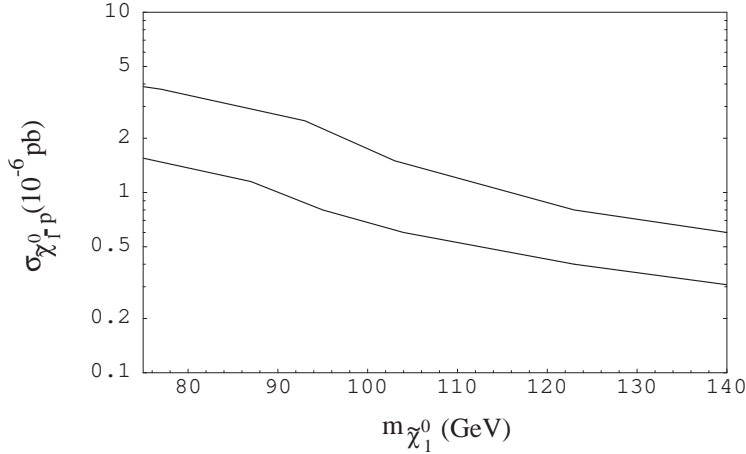

\centerline{ \DESepsf(aadtan712new.epsf width 10 cm) }
\caption {\label{fig7}Maximum value of $\sigma_{\tilde{\chi}_{1}^{0}-p}$ as a function of
$m_{\tilde{\chi}_{1}^{0}}$ for the nonuniversal
model with $\mu > 0$, $\delta_1$, $\delta_3$, $\delta_4 < 0$, $\delta_2 >0$. The lower curve
is for $\tan\beta = 7$, the upper curve is for $\tan\beta =12$.}
\end{figure}
To illustrate further some of the effects of nonuniversality, we consider
first the case where there are only Higgs mass nonuniversal soft breaking,
i.e. only $\delta_{1,2}$ are non zero. In general for the universal mSUGRA case,
both $\mu^2$ and $m_A$ are determined in terms of the other SUSY parameters.
From Eqs.~(\ref{eq28}) and (\ref{eq29}), we see that these quantities now depend on the
additional parameters $\delta_1$ and $\delta_2$, and thus may be varied keeping the
other (mSUGRA) parameters $m_0$, $m_{1/2}$, $A_0$ and $\tan\beta$ fixed. In
\cite{b40}, it was
thus assumed that $\mu^2$ and $m_A$ could be chosen arbitrarily. Actually,
however, both $\mu^2$ and $m_A$ are constrained by the bounds of
Eq.~(\ref{eq27}), and
varying $\mu^2$ and $m_A$ arbitrarily can lead to points in parameter space
with unreasonably large values of $\delta_{1,2}$ (or even tachyonic values with
$\delta_{1,2} < -1$). Thus here we will examine the actual parameter space
subject to the conditions of Eq.~(\ref{eq27}) (as well as the usual relic density
and accelerator bounds).
As an example of an effect of type (ii), we consider in Fig.8 the case in the
coannihilation region where $A_0=m_{1/2}$, $\tan\beta = 40$, and chose 
$\delta_2 = 1$.
Since $\mu^2$ is insensitive to $\delta_1$ we set $\delta_1 = 0$. In this domain, the
allowed stau coannihilation region corresponds to the lowest band in the $m_0-m_{1/2}$ plane
of Fig. 1, with $m_0$ running from about 200 GeV to 400 GeV and the
corresponding corridor is only slightly effected by $\delta_2\neq0$. As shown in Fig.
8, the effect of the nonuniversal soft breaking now allows the opening up of a
new wide band of allowed region at much higher values of $m_0$. Thus normally
such large values of $m_0$ would give rise to too little annihilation to
satisfy Eq.~(\ref{eq4}). However, by lowering $\mu^2$ and hence raising the higgsino content of the
neutralino, the $Z^0$ channel annihilation is
enhanced allowing for an acceptable relic density. In Fig. 8, the low side
of the band corresponds to $\Omega_{\tilde\chi^0_1} h^2 = $0.25, while the upper end to 
$\Omega_{\tilde\chi^0_1}h^2 =$0.02. If one were to raise the relic density 
lower bound to $\Omega_{\tilde\chi^0_1} h^2 = $0.1, the
width of the band would be reduced by about 30$\%$ to 40$\%$. (The small dip at
$m_{1/2} = (400-450)$ GeV (i.e. $m_{\tilde\chi^0_1} = $(160 - 180)GeV) arises due to the opening of the
$t-\bar t$ channel in the annihilation cross section.) Thus we see that
nonuniversal
soft breaking allows for much larger values of $m_0$ than mSUGRA and hence a
heavier mass spectrum, and this could be an experimental signal for
nonuniversalities. (Universal soft breaking only allows for the
coannihilation corridor in Fig. 8 at much lower $m_0$.) In addition, the
larger $m_0$ leads to a smaller $\sigma_{\tilde\chi^0_1-p}$.

\begin{figure}[htb]
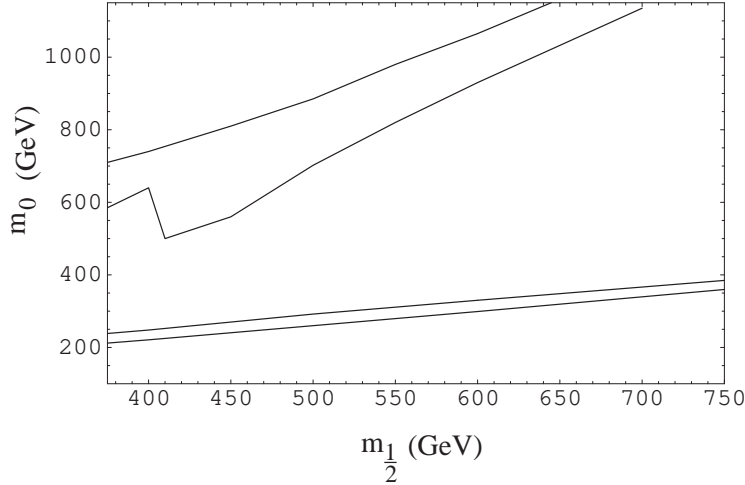

\centerline{ \DESepsf(aadcoan40newnon2.epsf width 10 cm) }
\caption {\label{fig8} Effect of a nonuniversal Higgs soft breaking mass enhancing the $Z^0$
s-channel pole contribution in the early universe annihilation, for the
case of $\delta_2 = $1, $\tan\beta = 40$, $A_0 = m_{1/2}$, $\mu > 0$. The lower band is
the usual $\tilde\tau_1$ coannihilation region. The upper band is an additional
region satisfying the relic density constraint arising from increased
annihilation via the $Z^0$ pole due to the decrease in $\mu^2$ increasing the
higgsino content of the neutralino.}
\end{figure}

We consider a second example where both effects of type (ii) and (iii)
occur simultaneously, by examining third generation squark and slepton
nonuniversal masses obtained when $\delta_{10} (=\delta_3=\delta_4=\delta_5)$ takes on
the value $\delta_{10} =$ -0.7. (All other $\delta_i$ are chosen zero.) We again
assume $A_0=m_{1/2}$, $\tan\beta = 40$ and $\mu >0$. The effect of $\delta_{10}$ is shown
in Fig. 9. The lowest band is the usual mSUGRA $\tilde\tau_1$ coannihilation band
(shown here for reference only). The upper band corresponds to two phenomena: The
lower half is the actual $\tilde\tau_1$ coannihilation region which has been shifted upwards
due to the fact that $m_0^2$ for the $\tilde\tau_1$ has been replaced by 
$m_0^2 (1 +\delta_{10}$), and since $\delta_{10}$ is negative, one needs a larger $m_0$ to achieve
the coannihilation corridor. The upper part corresponds to the opening of
the $Z^0$ channel as in Fig. 8 since $\delta_{10} < 0$ reduces $\mu^2$ increasing the
higgsino content of the neutralino. For $m_{1/2}\stackrel{<}{\sim}$ 500 GeV, the two bands
overlap. However at higher $m_{1/2}$, they are separated. Thus in the high $m_{1/2}$
region, the low side of the middle band (where $\tilde\tau_1$ coannihilation is
occuring) corresponds to $\Omega_{\tilde\chi^0_1} h^2 = $0.02, while the upper side corresponds
to $\Omega_{\tilde\chi^0_1} h^2 =$ 0.25. In the region between the two top bands, $\Omega_{\tilde\chi^0_1} h^2 >
$0.25 (and hence excluded). Finally, since $\delta_3$ and $\delta_4$ are negative
(reducing $\mu^2$ and also
making the total $m_0^2$ contribution to $\mu^2$ negative), for sufficiently
large $m_0$, the $Z^0$ channel opens reducing $\Omega_{\tilde\chi^0_1} h^2$ to 0.25 on the lower
side of the top band, that band terminating when $\Omega_{\tilde\chi^0_1} h^2 =$ 0.02 on the
top.
\begin{figure}[htb]
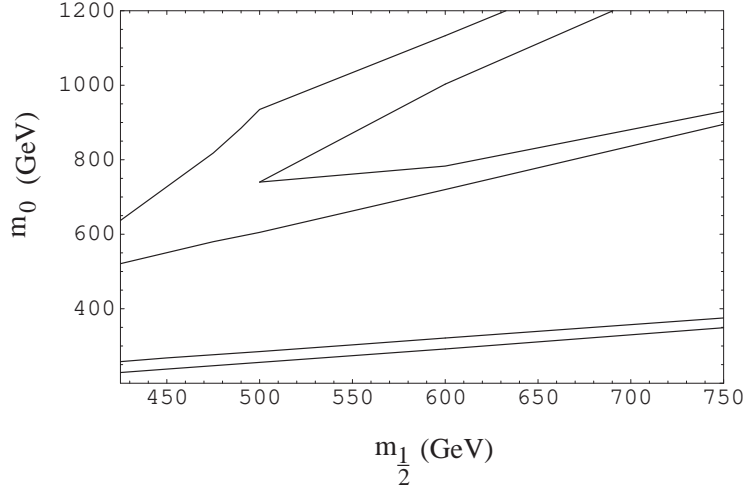

\centerline{ \DESepsf(aadcoan40newnon.epsf width 10 cm) }
\caption {\label{fig9} Allowed regions in the $m_0-m_{1/2}$ plane for the case 
$\tan\beta = 40$, $A_0
= m_{1/2}$, $\mu > 0$. The bottom curve is the mSUGRA $\tilde\tau_1$ coannihilation band of
Fig. 1 (shown for reference). The middle band is the actual $\tilde\tau_1$
coannihilation band when $\delta_{10} = -0.7$. The top band is an additional
allowed region due to the enhancement of the $Z^0$ s-channel annihilation
arising from the nonuniversality lowering the value of $\mu^2$ and hence raising the higgsino content of the neutralino. 
For $m_{1/2}\stackrel{<}{\sim}$ 500 GeV, the
two bands overlap.}
\end{figure}

We note in both Figs. 8 and 9, that the effect of nonuniversalities can
greatly enhance the allowed region in the $m_0-m_{1/2}$ plane for large $m_{1/2}$,
particularly by increasing $m_0$, and these changes occur without unreasonable
amounts of nonuniversality.

We consider next the neutralino-proton cross sections arising in these
nonuniversal soft breaking cases. One might have expected that the new
regions in the $m_0-m_{1/2}$ plane allowed by the relic density constraint would
correspond to smaller values of $\sigma_{\tilde\chi^0_1-p}$ since they are bands of much
larger $m_0$. However this is compensated by the fact that the nonuniversal
effects considered above also reduce $\mu^2$ and effects of type (i)
compensate for the largeness of $m_0$ and in fact give a net increase to the
cross section. This is illustrated for the example of Fig. 8 in Fig. 10
where $\sigma_{\tilde\chi^0_1-p}$ is plotted for $\delta_2 =$1, $\tan\beta =
40$, $A_0 =m_{1/2}$. The solid
line corresponds to the cross section for the $\tilde\tau_1$ coannihilation
corridor (the lower band of Fig.8), while the long (short)  dashed curves correspond to the cross
sections along the top (bottom) of the $Z^0$ pole band. We see that the $Z^0$
band cross sections, where $m_0$ is very large, are quite large, and this
region of parameters should soon be accesible to e.g. to the next phase of the
CDMS experiment in the Soudan mine. The cross section in the lower $m_0$
domain corresponding to the $\tilde\tau_1$ coannihilation region are much smaller,
and to explore this domain would require e.g. the GENIUS or Cryoarray
detectors.

In Fig. 11, we examine the expected neutralino-proton cross section for the
case of $\delta_{10} = -0.7$, $\tan\beta = 40$, $A_0 = m_{1/2}$, $\mu > 0$ of Fig. 9. Here the
large dashed curve is the cross section running along the top of the upper
band of Fig. 9 (the top of the $Z^0$ channel band) and the short dashed curve
is the cross section along the bottom of the middle band (the bottom of the
$\tilde\tau_1$ coannihilation corridor). The solid curve is shown for comparison and
would be the cross section in the $\tilde\tau_1$ coannillation corridor for universal
soft breaking (mSUGRA). We see again that the $Z^0$ corridor with very large
$m_0$ would be accesible to the CDMS Soudan experiment. The actual $\tilde\tau_1$
coannihilation corridor cross section lies considerably higher than the
universal case, but would probably require the GENIUS or Cryoarray
detectors to explore.

\begin{figure}[htb]
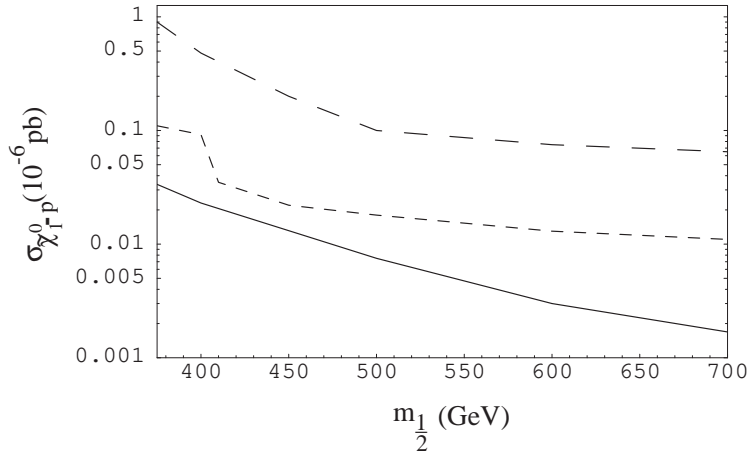

\centerline{ \DESepsf(aadcoan40newnon2cr.epsf width 10 cm) }
\caption {\label{fig10}$\sigma_{\tilde\chi^0_1-p}$ as a function of $m_{1/2}$ for $\delta_2 =
1$ (other $\delta_i = 0$),
for $\tan\beta=40$, $A_0 = m_{1/2}$, $\mu > 0$. The long (short) dashed curves are the
cross sections along the upper (lower) sides of the $Z^0$ bands of Fig. 8. The
solid curve is $\sigma_{\tilde\chi^0_1-p}$ along the $\tilde\tau_1$ coannihilation corridor of Fig. 8 .}
\end{figure}

\begin{figure}[htb]
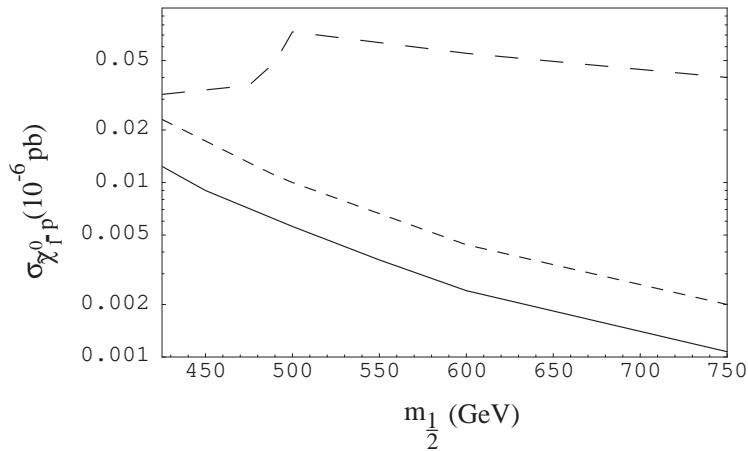

\centerline{ \DESepsf(aadcoan40newnon1cr.epsf width 10 cm) }
\caption {\label{fig11} $\sigma_{\tilde\chi^0_1-p}$ as a function of $m_{1/2}$
for $\delta_{10}$ = -0.7 (other 
$\delta_i$ =
0), for $\tan\beta = 40$, $A_0 = m_{1/2}$, $\mu > 0$. The long dashed curve is the cross
section along the top of the upper band in Fig. 9 where $Z^0$ annihilation is
occurring. (The reduction at $m_{1/2}\stackrel{<}{\sim} 500$ GeV is due to the 
$b\rightarrow s \gamma$
constraint.) The short dashed curve is the cross section along the bottom
of the middle band where $\tilde\tau_1$ coannihilation effects occur for this
nonuniversal model. The solid curve is included for comparison and is the
cross section in the $\tilde\tau_1$ coannihilation corridor for the case of universal
soft breaking (mSUGRA).}
\end{figure}

In the above discussion, we have chosen the signs of the $\delta_i$ parameters
to decrease $\mu^2$ to illustrate the dramatic effects such $\delta_i$ can
produce without any unusually large amount of nonuniversality. If instead
one were to chose the opposite signs, e.g. $\delta_2 < 0$, or $\delta_{10} > 0$, the
$\mu^2$ would have been increased. For this region of parameter space the $Z^0$
annihilation channel would now be absent (as in mSUGRA). The usual $\tilde\tau_1$
coannihilation corridors would of course still be present, but for the
$\delta_{10} > 0$ case at lower values of $m_0$. The cross sections for $\mu > 0$ would
then be somewhat reduced, but still accessible to future detectors.

The cross sections discussed above were for $\mu > 0$. We now consider the $\mu
< 0$ case. In the mSUGRA model, it was seen that for large $m_{1/2}$ with
$m_{1/2} < 1$ TeV and
a restricted range of $\tan\beta$, a special cancellation could occur which
reduced $\sigma_{\tilde\chi^0_1-p}$ to be below $10 ^{-12}$ pb (see Fig. 5), well below what
future detectors could see. This cancellation involved the Higgs $u$-quark
and Higgs $d$-quark amplitudes, when the neutralino components obey
Eq.~(\ref{eq24}).
However, nonuniversal choices such as those of Figs. (8-11), involve now
two region of parameter space consistent with the relic density bounds: the
coannihilation region, and the region above it dominated by the $s$-channel
$Z^0$ pole. Eq.~(\ref{eq24}) can in fact hold for the coannihilation corridors, and
the minima seen in the mSUGRA models occur here, at essentially the same
points as in Fig. 5. However, for the $Z^0$ bands, $\mu^2$ is quite reduced, and
$N_{14}$ is increased violating Eq.~(\ref{eq24}). Thus no such minima occur for $m_0$ in the $Z^0$
bands of Figs. 8 and 9.

\section{D-Brane Models}
Recent studies of Type IIB orientifolds have shown that it is possible to
construct a number of string models with interesting phenomenological
properties~\cite{b11}. The existance of open string sectors in Type IIB strings
implies the presence of Dp-branes, and in the simplest cases, one can
consider the full $D=10$ space compactified on a six torus $T^6$. Models of
this type contain either 9-branes and $5_i$-branes, $i=1,2,3$, or 3-branes and
$7_i$-branes. Associated with each set of n coincident branes is a gauge
group $U(n)$. There are a number of ways in which one can embed the Standard
Model in such systems. We consider here the model where $SU(3)_C\times U(1)_Y$ is
associated with one set of $5$-branes, $5_1$, and $SU(2)_L$ with a second
intersecting set $5_2$~\cite{b10}. Strings starting on $5_1$ and ending on $5_2$ then
carry the joint quantum numbers of the two sets of branes, the massless
modes then being the quark and lepton left doublets, and the two Higgs
doublets. Strings starting and ending on $5_1$ then carry the quantum numbers
of that brane, the massless modes being then the quark and lepton right
singlets.

Supersymmetry breaking can be treated phenomenologically by assuming that
the F components of the dilaton $S$ and the moduli $T_i$ grow
VEVs~\cite{b11,b41}:
\begin{eqnarray} 
F^S&=&{2\sqrt{3}}<{\rm Re} S>{\rm
sin}\theta_b e^{i\alpha_s}m_{3/2}\\ 
F^{T_i}&=&{2\sqrt{3}}<{\rm Re} T_i>{\rm
cos}\theta_b \Theta_i e^{i\alpha_i}m_{3/2}
\label{fsft}
\end{eqnarray}
where $\theta_b$, $\Theta_i$ are Goldstino angles ($\Theta_1^2 + \Theta_2^2 + 
\Theta_3^2
= 1$). In the following we assume $\Theta_3 = 0$. T-duality determines the
Kahler potential~\cite{b11,b41}, and using Eqs. (30,31) one generates the soft
breaking terms at $M_G$ for the gauginos,
\begin{eqnarray}
\tilde m_1&=&\tilde m_3=-A_0=\sqrt{3} \cos\theta_b \Theta_1
e^{-i\alpha_1}m_{3/2}\\
\tilde m_2&=&\sqrt{3} \cos\theta_b (1-\Theta_1^2)^{1/2}m_{3/2}
\end{eqnarray} and for the squarks, sleptons and Higgs,
\begin{eqnarray} 
m_{12}^2&=&(1-3/2 \sin^2\theta_b)m_{3/2}^2\, \,\,\,\, {\rm for} \,\,\,\,
q_L,\,l_L,\,H_1,\,H_2\\
m_{1}^2&=&(1-3 \sin^2\theta_b)m_{3/2}^2\,\,\,\,\, {\rm for}\,\,\,\,
u_R,\,d_R,\, e_R.
\end{eqnarray} 
Originally this model was considered to examine the possible effects of the
new CP violating phases in Eqs.(32,33) and in the $\mu$ and $B_0$
parameters~\cite{b10,b42}. However, it was subsequently seen that unless $\tan\beta$ is
small (i.e. $\tan\beta\stackrel{<}{\sim}5$) an unreasonable fine tuning of the GUT
scale parameters is needed if the experimental constraints on the electron and
neutron electric dipole moments are to be obeyed~\cite{b42}. The LEP data has now
eliminated a considerable amount of the low $\tan\beta$ region, and since we
are interested here in effects that arise for large $\tan\beta$, we will in the
following set all the phases to zero. The model then depends on three
parameters: $\theta_b$, $\Theta_1$, and $m_{3/2}$, with $\theta_b < 0.615$ and 
$0 < \Theta_1<1$, and contains a unique type of nonuniversal soft breaking. In terms of
the notation of Eq. (26) one has
\begin{eqnarray}
          \delta_1&=&\delta_2= \delta_3= \delta_7 = -(3/2) \sin^2\theta_b\\
	   \delta_4 &=& \delta_5 = \delta_6 = -3 \sin^2\theta_b
\end{eqnarray}
where we have equated $m_0$ to $m_{3/2}$. Note that these nonuniversalities hold
for all three generations, not just the third generation. In addition the
gaugino masses at $M_G$ are no longer universal (except when $\Theta_1
=1/\sqrt{2}$), and there are two ``effective " gaugino masses at $M_G$ which we
label by $m_{1/2} = \tilde m_1 = \tilde m_3$, and $m_{1/2}' = \tilde m_2$. 
The $A_0$ parameter in this model
is then fixed by $A_0 = - m_{1/2}$.

We consider first the question of what parts of the parameter space might
be accessible to current detectors which obey Eq. (1). From Eqs. (34,35) we
see that the sfermion masses decrease with increasing $\theta_b$, and so we
expect the largest cross sections to arise for small $\theta_b$, and of course
for large $\tan\beta$. However, the $b \rightarrow s \gamma$ constraint
eliminates the higher values of $\theta_b$. The limiting boundary obeying Eq. (1) is shown in Fig.
12, where $\sigma_{\tilde\chi^0_1-p}$ is plotted for $\tan\beta = 20$ and $\theta_b = 0.2$. For this
case, the neutralino mass would be quite small but is still consistent with
the LEP bounds\footnote{LEP determines the bound on 
$m_{\tilde\chi^0_1}$ by measuring lower bounds on the chargino
mass and lower bounds on $m_{\tilde\chi^0_1}+m_{\tilde\chi^0_2}$~\cite{b43}. This combined with lower
bounds on the Higgs and slepton masses and the value of the $Z$ width, limits
the SUSY parameter space giving rise to the $m_{\tilde\chi^0_1}$ bound. 
Thus the bounds depend
on the SUSY model used, and LEP quotes 37 GeV for the MSSM, and 48 GeV for
mSUGRA. We have checked that for the D-brane model being considered, the
LEP data still implies $m_{\tilde\chi^0_1} > 37$ GeV.
}. (The gap in the middle is due to over annihilation in the early universe through
the s-channel $Z$ pole.) The cross section rises, of
course with higher $\tan\beta$, and so current detectors are sampling regions
where
\begin{equation}
 \tan\beta\stackrel{>}{\sim} 25 
\end{equation}

It is easy to see that slepton coannihilation effects exist in this model
similar to those of the SUGRA models. Since both the $\tilde e_R$ and 
$\tilde\chi^0_1$ both
evolve from the GUT scale using the $U(1)$ gaugino, Eq.(16) still holds with
$m_0^2$ replaced $m_{3/2}^2 (1 - 3\sin^2\theta_b)$. Thus for fixed $\theta_b$, one can
adjust $m_{3/2}$ so that the $\tilde\chi^0_1$ and $\tilde e_R$ are nearly degenerate, allowing for
coannihilation to take place. In addition, however, there can be regions of
parameter space where there is chargino-neutralino coannihilation. This arises because of the
nonuniversal gaugino masses at $M_G$. Thus from Eqs. (32,33), when $\Theta_1$
gets large, $\tilde m_2$ becomes smaller than $\tilde m_1$, and the $\tilde\chi^0_1$
 becomes mainly
Wino, and can become degenerate with the $\tilde\chi^{\pm}_1$. Coannihilation becomes
possible, however, only for a very small range in $\Theta_1$ near 0.9.

\begin{figure}[htb]
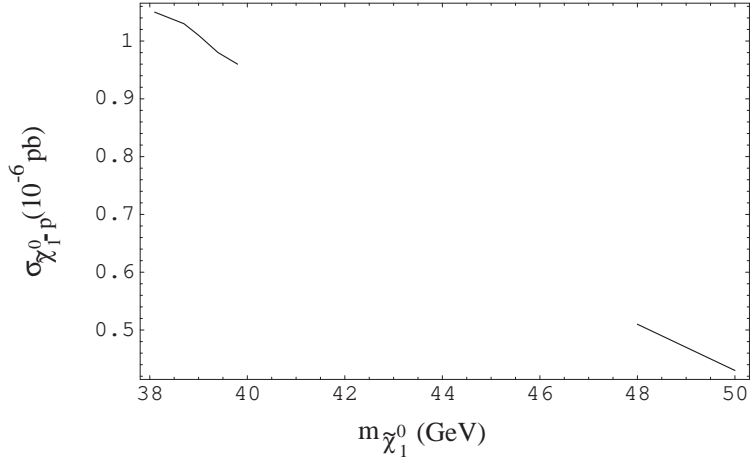

\centerline{ \DESepsf(aadtan20dark.epsf width 10 cm) }
\caption {\label{fig12} $\sigma_{\tilde\chi^0_1-p}$ for D-brane model for $\mu > 0$, 
$\theta_b=0.2$ and $\tan\beta = 20$. The gap in the curve is due to excessive early universe
annihilation through the s-channel $Z$ pole.}
\end{figure}

As in SUGRA models, a cancellation of matrix elements can occur for $\mu < 0$,
allowing the cross sections to fall below the sensitivities of planned
future detectors. This is exhibited in Fig. 13, where
$\sigma_{\tilde{\chi}_1^0-p}$ is plotted for $\tan \beta = 6$ (solid curve), 12
(dot-dash curve), and 20 (dashed curve). (The $\tan \beta = 6$ curve terminates
at low $m_{\tilde{\chi}_1^0}$ due to the $m_h$ constraint, while the higher $\tan
\beta$ curves terminate at low $m_{\tilde{\chi}_1^0}$ due to the $b \rightarrow s
\gamma$ constraint. The upper bound on $m_{\tilde{\chi}_1^0}$, corresponding to
$m_{\tilde{g}} < 1$ TeV, arises from the $\Omega h^2$ constraint.) One sees that
the cross section goes through a minimum at $\tan \beta \cong 12$, though the
expected rise at higher $m_{\tilde{\chi}_1^0}$ does not appear since the
parameter space terminates before this sets in.

\begin{figure}[htb]
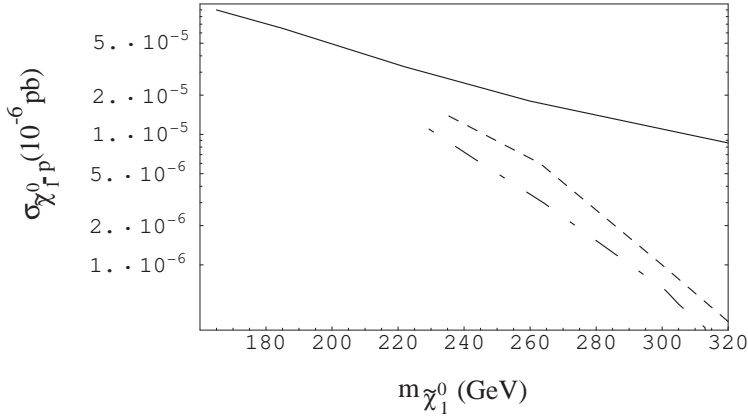

\centerline{ \DESepsf(adposmudark61215.epsf width 10 cm) }
\caption {\label{fig13} $\sigma_{\tilde\chi^0_1-p}$ for the D-brane model for $\mu < 0$ and
$\tan\beta = 6$ (solid), $\tan\beta = 12$ (dot-dash), and $\tan\beta = 15$ (dashed).}
\end{figure}

\section{Stop And Sbottom Coannihilation.}

It has been known for some time that the nearness of the $t$-quark Landau
pole can significantly lower the $\tilde{t}_1$ mass~\cite{b44}, and this raises the
possibility of $\tilde{t}_1-\chi$ coannihilation occurring~\cite{c3}. Working against this
possibility is the positive contributions to $m_{\tilde{t}_1}^2$ from the
$SU(3)$
gaugino contribution, and of course the $m_t^2$ contribution. Nonuniversal
soft breaking, however, opens additional possibilities of this type, as it
can allow the coefficient of $m_0^2$ to become negative, reducing further the
$\tilde{t}_1$ mass. To examine this in more detail, we consider the low and
intermediate $\tan\beta$ regime where the RGE can be solved
analytically. For
the $\tilde{t}_R$ one finds
\begin{eqnarray}
 m_{\tilde{t}_R}^2 &=& \left[ D_0 +(-\frac{1-D_0}{3}(\delta_2+\delta_3)+
 \frac{2+D_0}{3} \delta_4 ) \right] m_0^2 \nonumber \\
 &+& \frac{2}{3} \left[ (1-D_0)D_0 \frac{H_3}{F} A_0 m_{1/2} - \frac{1}{2}
 (1-D_0)D_0 A_0^2 + e m_{1/2}^2 \right] \nonumber \\
 &+& (\frac{1}{3} f_1 - f_2 + \frac{8}{3} f_3 ) \frac{\alpha_G}{4 \pi} m_{1/2}^2 +
 m_t^2 + \frac{2}{3} \sin^2 \theta_W M_Z^2 \cos 2 \beta
\end{eqnarray}                                                                  
The $f_i$ form factors, $i=1,2,3$, are numerically approximately 40, 100, 700
respectively, and $e$, and $H_3/F$ given in~\cite{b45}, are approximately
-2 and +2 respectively. One sees that for $A_0$ large and negative, the Landau pole $A_0$
terms do give a significant negative contribution and $m_{\tilde{t}_R}^2$ would
be minimized for $\delta_{2,3} > 0$ and $\delta_4 < 0$. The universal $m_0^2$
term, scaled by $D_0 \cong 0.25$, is already quite small, so it does not require
too much
nonuniversality to achieve a degeneracy with the $\chi_1^0$. (In fact this can
be done even for the universal case if $A_0$ and $m_{1/2}$ are chosen
sufficiently
large. However, the degeneracy occurs in that case for parameters in the
TeV domain.) Since early universe annihilation cross sections involving the
stop are strong interaction phenomena compared with the electroweak
neutralino cross sections, the coannihilation effect can be quite
significant. However, it is necessary to impose the usual constraints,
particularly the $b \rightarrow s  \gamma$ bounds, and also that the stop must lie
below
the stau. Thus a satisfactory coannihilation region occurs only when $m_0$ is
large (e.g. greater than about 800 GeV). The neutralino-proton cross section
in this domain generally lies above $10^{-10}$ pb, and thus should be
accessible to future detectors.

Stop coannihilation phenomena are more subtle in the D-brane model, since
the nonuniversalities are related i.e. $\delta_2 = \delta_3 =
-(3/2)\sin^2\theta_b m_{3/2}^2 = \delta_4/2$. The $m_0^2$ term for this case
is
\begin{equation}
                 [D_0 -(1 + 2D_0)\sin^2\theta_b]m_{3/2}^2                          
\end{equation}
which is negative for $\sin^2\theta_b \stackrel{>}{\sim} 1/6$. (There are also
modifications in
the gaugino terms to acount for their nonuniversal nature in the D-brane
model.)  Thus stop coannihilation effects can occur also in this model.
Here, however, the parameter space for coannihilation is narrowed and
generally occurs when $m_{3/2} > 1$ TeV.

$b$-squark coannihilation phenomena can also occur in these models. For low
and intermediate $\tan\beta$, where one may neglect the $b$-quark Yukawa
coupling
(relative to the $t$-quark coupling), the $\tilde{b}_R$ mass is generally large
(due to the $SU(3)$ gluino contribution). However, the $\tilde{b}_L$ mass can
get small for
nonuniversal soft breaking. Thus $m_{\tilde{b}_L}$ is given by
\begin{eqnarray}
  m_{\tilde{b}_L}^2 &=& \left[ \frac{1+D_0}{2} +(- \frac{1-D_0}{6} (\delta_2 +
  \delta_4) + \frac{5+D_0}{6} \delta_3) \right] m_0^2 \nonumber \\
  &+& \frac{1}{3} \left[ (1-D_0) D_0 \frac{H_3}{F} A_0 m_{1/2} - \frac{1}{2}
  (1-D_0)D_0 A_0^2+e m_{1/2}^2 \right] \nonumber \\
  &+& (-\frac{1}{15} f_1 + f_2 + \frac{8}{3} f_3 ) \frac{\alpha_G}{4 \pi}
  m_{1/2}^2 + (-\frac{1}{2}+\frac{1}{3} \sin^2 \theta_W) M_Z^2 \cos 2 \beta +
  m_b^2 
\end{eqnarray}
and the light $b$-squark can be achieved by chosing $\delta_2, \delta_4 >0$ and
$\delta_3 <0$ so that the coefficent of the $m_0^2$ term is negative. In this
case
the light $b$-squark is mostly $\tilde{b}_L$, and since the $t$ and $b$ squarks
form an $SU(2)$ doublet, the light stop lies just above the the light sbottom,
and there can be regions where both contribute simultaneously to
coannihilation. However, these coannihilation regions again occur for $m_0 >
1$ TeV.

\section{Conclusions}

We have considered in this paper coannihilation effects in dark matter
neutalino-proton detection cross sections which arise for large $\tan\beta$ and
large values of $A_0$ for models which have grand unification of the gauge
coupling constants at the GUT scale $M_G$. Coannihilation can occur if the light
stau ($\tilde{\tau}_1$), light chargino ($\tilde{\chi}^{\pm}_1$), light stop
($\tilde{t}_1$) or light sbottom
($\tilde{b}_1$) become nearly degenerate with the neutralino. In order to
examine
these effects for a wide range of models, we have considered SUGRA models
with universal soft breaking (mSUGRA), nonuniversal soft breaking in the
Higgs and third generation squarks and sleptons, and finally a D-brane
model based on type IIB orientifolds which possess nonuniversal gaugino
masses. In all these we have imposed the cosmological neutralino relic
density constraints as well as accelerator constraints on the SUSY
parameter space (including in particular the large $\tan\beta$ NLO corrections
to the $b \rightarrow s\gamma$ decay~\cite{b16,b17} as well as the large
$\tan\beta$ SUSY
corrections to $m_b$ and $m_{\tau}$). Radiative breaking of $SU(2) \times U(1)$
at the
electroweak weak scale also gives strong constraints on the theory.

mSUGRA illustrates some of the effects arising for large $\tan \beta$. Thus for
low $\tan \beta$, the regions in the $m_0 - m_{1/2}$ plane allowed by the relic
density constraint Eq. (\ref{eq4}) is relatively insensitive to $A_0$. However,
for large $\tan \beta$, the $b \rightarrow s \gamma$ constraint allows only the
$\tilde{\tau}_1 - \tilde{\chi}_1^0$ coannihilation corridors to remain (except
for some small islands when $m_{1/2} \stackrel{<}{\sim} 350$ GeV) and as $A_0$
increases, these corridors have increasing $m_0$. Thus $m_0$ can rise to 1 TeV
for large $m_{1/2}$. (See e.g. Fig. 1.) This lowers the value of
$\sigma_{\tilde{\chi}_1^0-p}$ and even though $\sigma_{\tilde{\chi}_1^0-p}$
increases rapidly with $\tan \beta$, the value of $\sigma_{\tilde{\chi}_1^0-p}$
at large $\tan \beta$ and large $A_0$ can be smaller than
$\sigma_{\tilde{\chi}_1^0-p}$ at $\tan \beta = 3$, as illustrated in Fig.4. For
$\mu > 0$, however, one generally finds for the parameter space of Eqs. (6-8)
that  $\sigma_{\tilde{\chi}_1^0-p} \stackrel{>}{\sim} 10^{-10}$ pb, and so
should be accessible to future detectors. In the high $\tan \beta$ domain, the
$b \rightarrow s \gamma$ constraint produces a lower bound on $m_{1/2}$, which
decreases with increasing $A_0$. Thus one finds that $m_{\tilde{\chi}_1^0}
\stackrel{>}{\sim} 120$ GeV ($m_{\tilde{g}} \stackrel{>}{\sim} 740$ GeV) for
$A_0 = 4 m_{1/2}$, $\tan \beta = 40$, $\mu >0$.

For $\mu < 0$, a special cancellation can occur in mSUGRA allowing
$\sigma_{\tilde{\chi}_1^0-p}$ to become much less than $10^{-10}$ pb, as has
been previously noted for low and intermediate $\tan \beta$~\cite{b30}. The
analytic origin of this effect is discussed in Sec. 2. The effect, however,
occurs only for a fixed range of $\tan \beta$. Thus the position of these minima
begins for $\tan \beta \cong 7$ at $m_{1/2} = 1000$ GeV, moves to lower
$m_{1/2}$ until $\tan \beta$ increases to 10, and then increases to beyond
$m_{1/2} = 1000$ GeV at $\tan \beta \stackrel{>}{\sim} 30$. (See Fig. 5.) We
note that while the cancellation occurs at a fixed value of $m_{1/2}$, the
effects spread over a wide range of $m_{1/2}$ and in this region, halo dark
matter would be inaccessible to current or planned detectors. However, if the
recently reported anomaly in the muon gyromagnetic ratio~\cite{b46} is indeed
due to
supersymmetry, then $\mu$ is positive and this case would be eliminated.

The nonuniversal soft breaking models illustrate new phenomena at large $\tan
\beta$ not present in mSUGRA, and one can also understand these effects
analytically. Thus the value of $\mu^2$, obtained from radiative breaking of
$SU(2) \times U(1)$, is a major element in controlling the value of
$\sigma_{\tilde{\chi}_1^0-p}$, and lowering $\mu^2$ increases
$\sigma_{\tilde{\chi}_1^0-p}$. Thus for one sign of the nonuniversalities,
$\sigma_{\tilde{\chi}_1^0-p}$ can be increased by a factor of 10 or more,
allowing current detectors to sample the parameter space with $\tan \beta$ as
low as $\cong 7$. (See Fig. 7.) For large $\tan \beta$ a new effect can occur.
The value of $\mu^2$ also controls the gaugino/higgsino content of the
$\tilde{\chi}_1^0$, and lowering $\mu^2$ increases the amount of Higgsino in the
$\tilde{\chi}_1^0$. This then increases the amount of early universe
annihilation via an off shell s-channel $Z$ boson, which opens an additional
allowed region in the $m_0-m_{1/2}$ plane, considerably wider than the
$\tilde{\tau}_1$ coannihilation corridor and at much higher $m_0$. (See Fig. 8.)
(Other simple choices of nonuniversal parameters can also raise the values of
$m_0$ for the  $\tilde{\tau}_1$ coannihilation corridor, as illustrated in
Fig.9.) Though $m_0$ is large for this $Z$-channel annihilation, reducing
$\sigma_{\tilde{\chi}_1^0-p}$, this effect is compensated by the fact that
$\mu^2$ is decreased (which raises $\sigma_{\tilde{\chi}_1^0-p}$). Thus
$\sigma_{\tilde{\chi}_1^0-p}$ remains relatively large (see Figs. 10, 11), and
this region of
parameter space should become experimentally accessible in the relatively near
future (e.g. when CDMS moves to the Soudan mine).

The D-brane model examined is quite constrained as all the squark and slepton
nonuniversalities are controlled by a single parameter $\theta_b$. Current
detectors are sampling parts of the parameter space where $\tan \beta
\stackrel{>}{\sim} 25$. Because these models have also nonuniversal gaugino
masses, the LEP lower bound on $m_{\tilde{\chi}_1^0}$ is much smaller than in
SUGRA models, i.e. the same as in the MSSM: $m_{\tilde{\chi}_1^0} \geq 37$ GeV.
However, the gaugino nonuniversality still allows the usual type of
$\tilde{\tau}_1-\tilde{\chi}_1^0$ coannihilation to occur. A unique feature of
the gaugino nonuniversality is that $\tilde{m}_2$ can become smaller than
$\tilde{m}_1$ allowing the $\tilde{\chi}_1^0$ to become mostly Wino and a
$\tilde{\chi}_1^{\pm}-\tilde{\chi}_1^0$ coannihilation domain to exist. However,
the near degeneracy of $\tilde{\chi}_1^{\pm}$ and $\tilde{\chi}_1^0$ occurs only
for a very small region of parameter space (when $\Theta_1 \cong 0.9$). The
cancellation phenomena in $\sigma_{\tilde{\chi}_1^0-p}$ for $\mu < 0$ also
occurs for these D-brane models, the minimum value of $m_{\tilde{\chi}_1^0}$
where the cancellation is complete occuring at $m_{\tilde{\chi}_1^0} \cong 315$
GeV for $\tan \beta \cong 12$.

Nonuniversal SUGRA models also allow for $\tilde{t}_1-\tilde{\chi}_1^0$
coannihilation to occur. In this case, the light stop is mostly $\tilde{t}_R$.
Then for $A_0 < 0$, the Landau pole contributions to $m_{\tilde{t}_R}^2$ are
negative, and one can chose nonuniversal soft breaking so that the coefficient
of $m_0^2$ is also negative. The $\tilde{t}_1$ then becomes nearly degenerate
with the $\tilde{\chi}_1^0$ when $m_0 \stackrel{>}{\sim} 800$ GeV,
coannihilation setting in for that domain. Actually,
$\tilde{t}_1-\tilde{\chi}_1^0$ can also occur in mSUGRA, but to achieve the near
degeneracy one needs $A_0 > 4 m_{1/2}$. Similarly, this coannihilation can occur
for the D-brane model but for $m_{3/2} > 1$ TeV.

In a similar fashion, $\tilde{b}_1-\tilde{\chi}_1^0$ coannihilation can occur
for nonuniversal models when the $\tilde{b}_1$ is mainly $\tilde{b}_L$. Thus one
may chose $A_0 < 0$ and nonuniversal parameters in $m_{\tilde{b}_L}^2$ so that
the coefficient of $m_0^2$ is negative, making the $\tilde{b}_1$ nearly
degenerate with the $\tilde{\chi}_1^0$. This coannihilation only sets in,
however, for $m_0 > 1$ TeV.

\section{Acknowledgement}
This work was supported in part by National Science Foundation grant No.
PHY-0070964. 

\bigskip
\noindent
Note added: After completing this paper there appeared the paper ``The CMSSM
Parameter Space at Large $\tan \beta$'' by J. Ellis, T. Falk, G. Ganis, K.A.
Olive and M. Srednicki, hep-ph/0102098. This paper considers large $\tan \beta$
effects for the mSUGRA model only, and is restricted to $A_0=0$.

\end{document}